\documentclass{article}

\usepackage{arxiv}

\usepackage[utf8]{inputenc} 
\usepackage[T1]{fontenc}    
\usepackage{hyperref}       
\usepackage{url}            
\usepackage{booktabs}       
\usepackage{amsfonts}       
\usepackage{nicefrac}       
\usepackage{microtype}      
\usepackage{lipsum}
\usepackage{wrapfig}
\usepackage{graphicx}
\usepackage[export]{adjustbox}
\usepackage{subcaption}
\usepackage{caption}
\usepackage{multicol}
\usepackage{colortbl}
\usepackage[dvipsnames]{xcolor}
\usepackage{tikz}
\usepackage{float}
\usepackage{amsmath}

\title{Device for ECG prediction based on retinal vasculature analysis}

\author{
Pranjal Rai\\
ee1180484@iitd.ac.in\\
\And
Sachin Jangir\\
ee1180495@iitd.ac.in\\
\And
Tanvir Singh Bal\\
ee1180508@iitd.ac.in\\
}

\begin{document}
\maketitle


\vspace{-0.5cm}
\section{INTRODUCTION}
Pulsatile changes in retinal vascular geometry over the cardiac cycle
have clinical implications for the diagnosis of ocular and systemic
vascular diseases, including ischemia, coronary heart diseases,
and diabetes mellitus and its complications. Thus, analysis of the
pulsatile changes over the cardiac cycle is a potential non-invasive
assessment for the presence of ocular and systemic vascular diseases. 
The cardiac rhythm influences these pulsatile changes in the retina, is
a result of the change in blood volumetric flow entering the ophthalmic-vascular system under a certain level of intraocular pressure during the
peak systolic and diastolic phases of the cardiac cycle.
An example of pulsatile change observable in the retina is the
spontaneous venous pulsation (SVP), which is available in
approximately 90 percent of the patients. It is caused by the
variation in the pressure gradient between the intraocular retinal
veins and the retrolaminar portion of the central retinal vein, visible
as rhythmic changes in the diameter of one or more veins near or
on the optic nerve head. In addition to SVP, pulsation of veins
outside the optic disk (OD), such as the serpentine movement of
principal arteries, the pulsatile motion of small arterioles, and movement
of optic nerve head are other features that can be visualized with dynamic fundoscopy.

Assessment of fundus images generally requires ophthalmologic expertise. However, 
the availability of an expert is not always guaranteed, and even if an expert is available, the assessment is performed manually.
Thus, there is also a need for an automated device that can
analyze the fundus images and keep track of the pulsations.
Such a device can be synchronized with the Electrocardiogram
and can be programmed to assess the presence of various ocular and
systemic vascular diseases. In this project, we proposed and worked on a portable embedded system device that automatically captures retinal images and analyzes them for the estimation of vessel diameters. The device is hand-held and attaches to the ophthalmoscope as an extension. Using the device, the time variation of the vessel diameters can be estimated, which can further be used to predict ECG and the presence of various other systemic vascular diseases.

\section{STRUCTURE OF THE DEVICE}
\label{sec:structure}
The device is integrable with an ophthalmoscope and acts as an extension to the
ophthalmoscope. The functions performed by it are to capture and sample
the retinal images, process them for vessel diameter estimation and output
the time variation of those estimations, which are to be used further for ECG prediction.

\subsection{COMPONENTS AND THEIR FUNCTIONS}
A Raspberry Pi processor was used as the processing part of the device, 
whose function is to initiate the image capturing and process the images 
for vessel diameter estimation. 
The main sensing element of the device is the camera, and the Raspberry Pi camera module provided a Sony IMX219 8-megapixel sensor for the device which can support 1080p 30, 720p 60, and VGA 90 video modes.
The device also needed an input-output tool, and a 3.5 inch LCD screen was chosen to be pertinent with the device.
Apart from the above main parts, various secondary parts like connecting battery, connecting wires, and screws were used.
Details of the components are also summarised in Table \ref{tab:components}.
All the components mentioned above work hand-in-hand. 

\begin{figure}[h]
\centering
\includegraphics[width=0.8\linewidth]{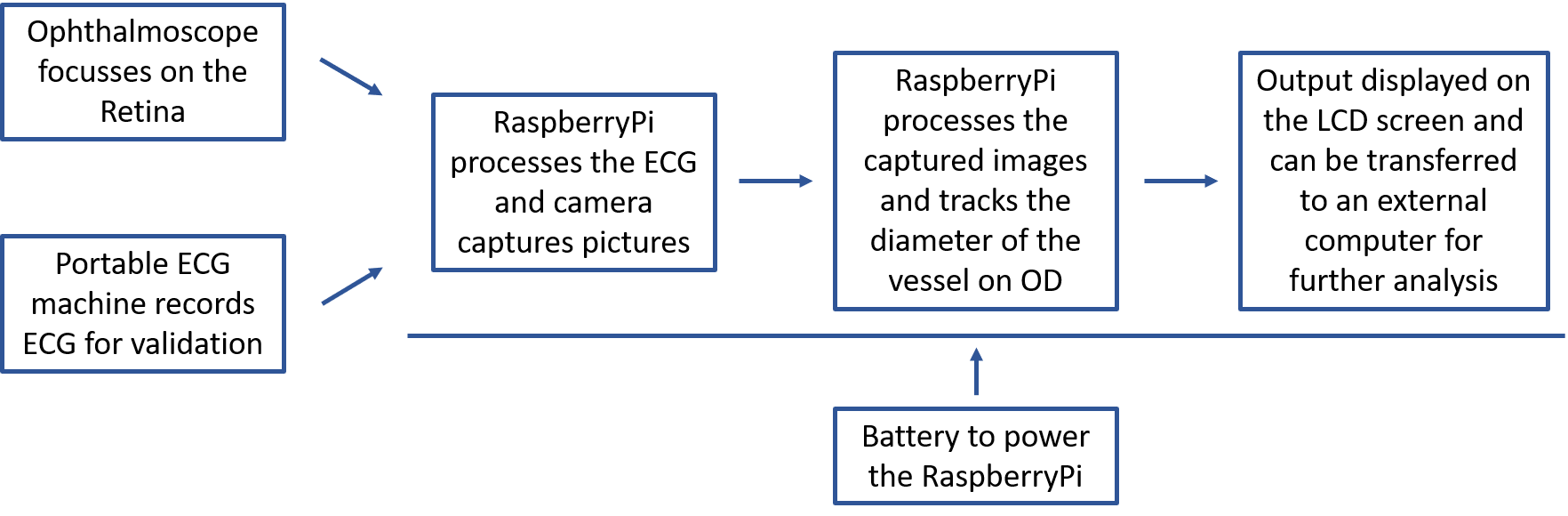}
\caption{Block diagram for the device}
\label{fig:block_diagram}
\end{figure}

The ophthalmoscope focuses on the retina, and the Raspberry Pi camera captures the retina's focussed image. The images are passed on to a Raspberry Pi powered by a battery.
The Raspberry Pi processes the images and tracks the time variations in the vessel diameters. The processed output is passed on to the user via the LCD screen.
The device can also be used synchronized with the ECG to visualize retinal pulsations. For this, 
a portable ECG machine is used to continuously capture the ECG and passes it to the Raspberry Pi. The processor
uses those measurements to detect peaks and trigger image capture at the ECG's specified peaks. The images are
then processed to obtain the time variation of vessel diameters, which can be used to study the pulsations.
The block diagram is shown in Figure \ref{fig:block_diagram} 

\begin{table}[h]
\caption{Components of the device}
\centering
\begin{tabular}{lll} 
\cmidrule(r){1-2}
Component     & Status     \\
\midrule
Ophthalmoscope & Present in the lab\\
Raspberry Pi     & Issued from the lab\\
Raspberry Pi camera module     & Issued from the lab\\
3.5 inch LCD screen     & Rs. 3000\\
3.4V 4000mAh Li-ion battery     & Rs. 600\\
Misc. (wires, screws, etc.)     & Rs. 200\\
\bottomrule
\end{tabular}
\label{tab:components}
\end{table}
\vspace{-0.5cm}

\subsection{3D MODEL}
\begin{figure}[H]
\centering
\includegraphics[width=0.8\linewidth]{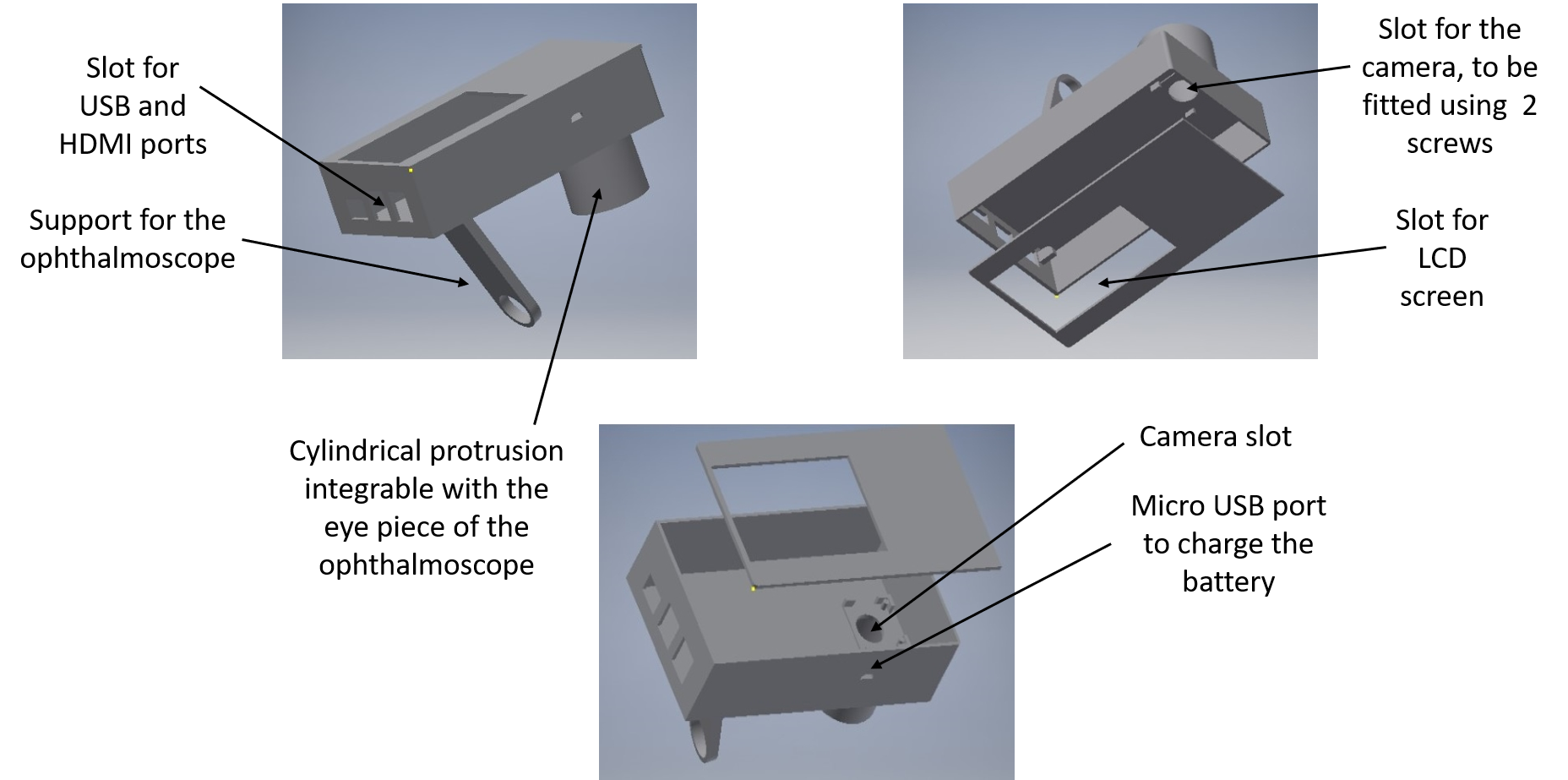}
\caption{3D model of the device}
\label{fig:model}
\end{figure}

\begin{figure}[H]
\centering
\includegraphics[width=0.8\linewidth]{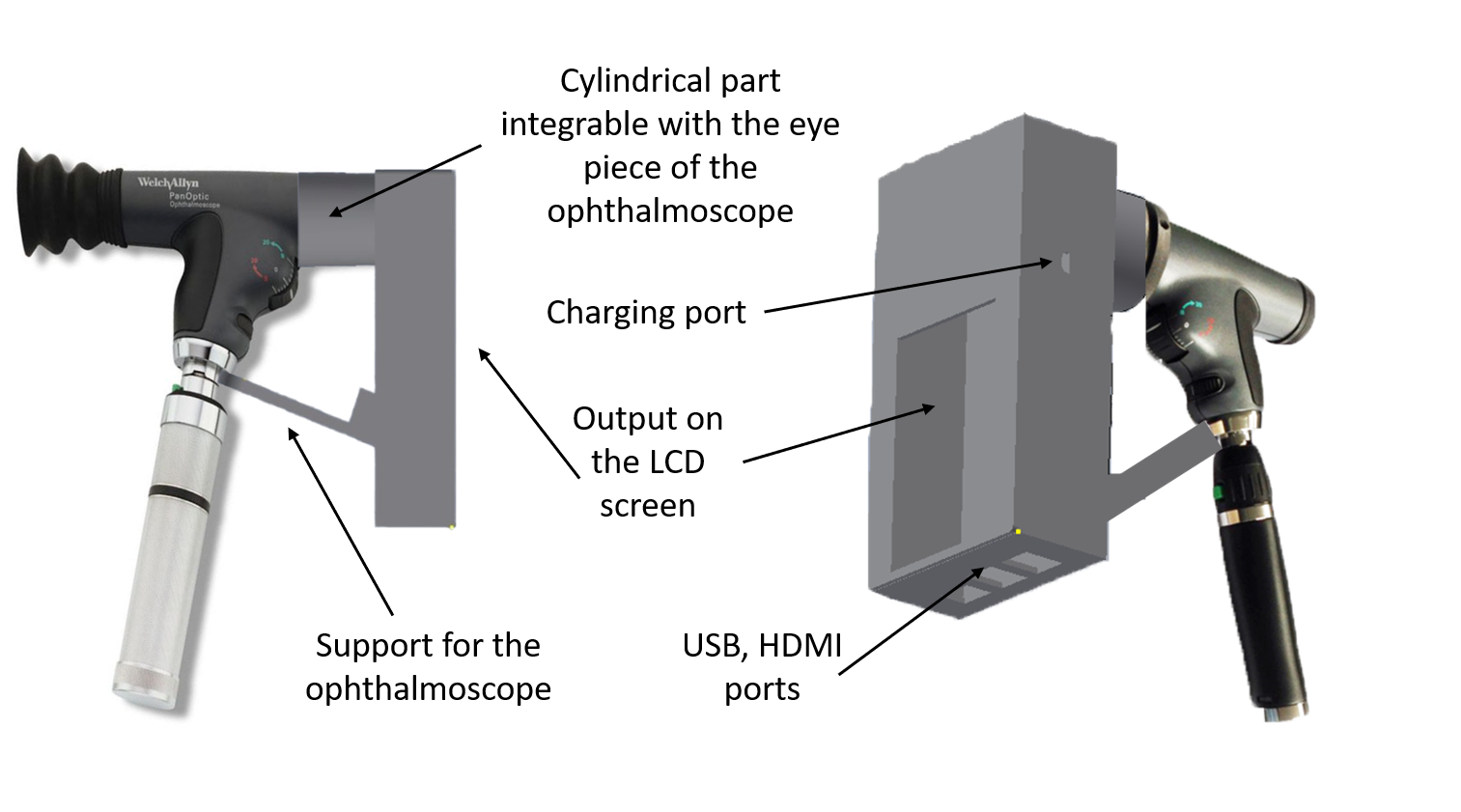}
\caption{Device integrated with an ophthalmoscope}
\label{fig:iso}
\end{figure}

\begin{figure}[H]
\centering
\includegraphics[width=0.8\linewidth]{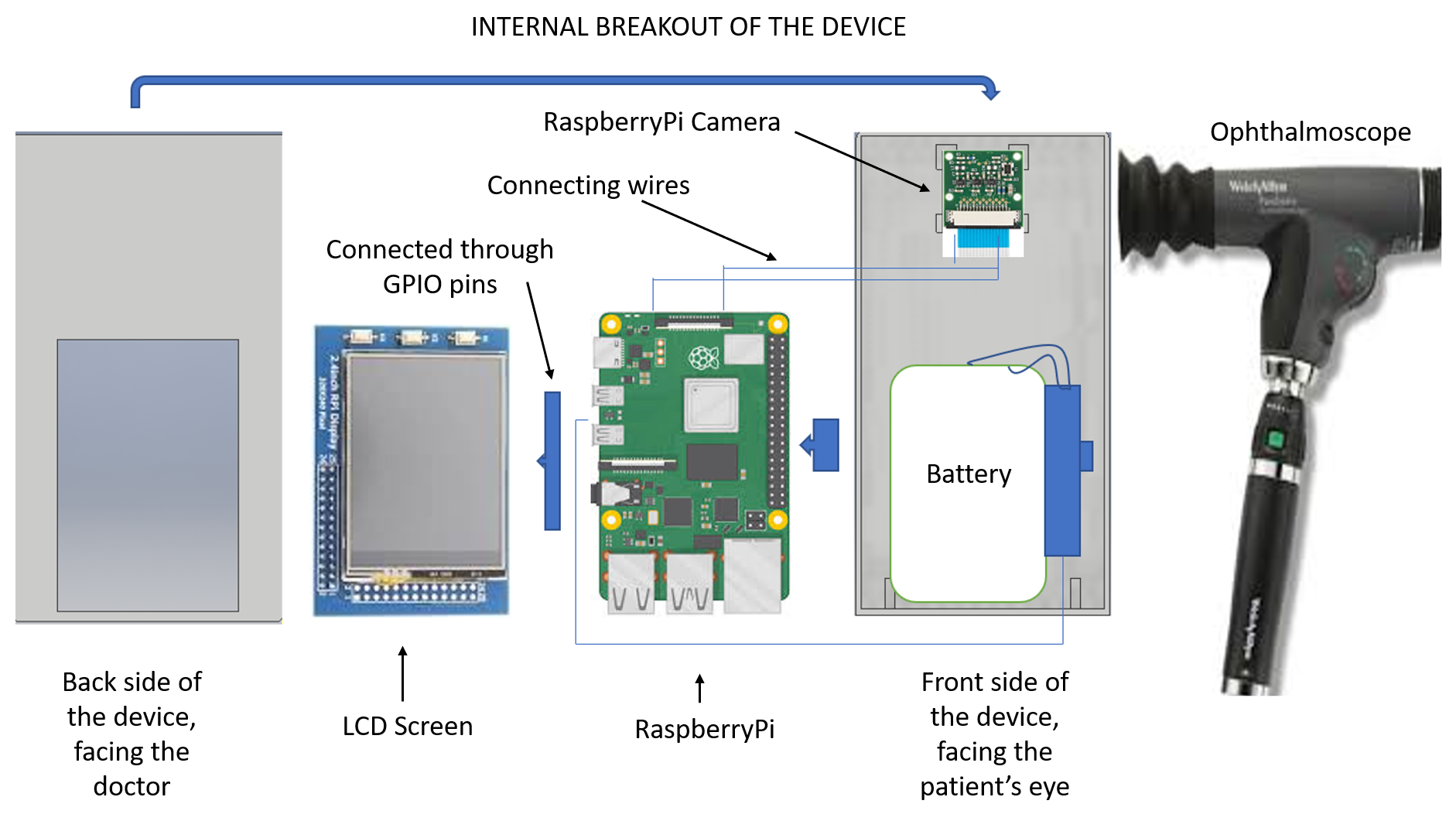}
\caption{Internal breakout of the device}
\label{fig:internal}
\end{figure}

All the components will function only when there is harmony in their physical
arrangement, and that is where modeling the device comes into the picture. The 3D
model of the device was made using the Autodesk Inventor software. The model was designed
to provide space for every component and the connections between the components. Ports like 
USB, HDMI, and micro USB were incorporated into the model. A unique cylindrical
protrusion was designed to attach to the eye-piece of the ophthalmoscope and
thus provide a pseudo-continuum between the ophthalmoscope and the device. The stability of the
model was also taken into consideration, and support was designed to provide the device robustness when
connected with an ophthalmoscope.
The 3D model and the internal breakout are shown in the Figures \ref{fig:model}, \ref{fig:iso} and \ref{fig:internal}.

\section{WORKING OF THE DEVICE}
The Raspberry Pi triggers image capture when input is given through the 
LCD screen. The captured image is processed for the estimation of vessel diameters.
The state of the art algorithms for diameter estimation generally use Convolutional
Neural Networks. The dataset these algorithms are trained on is DRIVE, STARE, and REVIEW.
These datasets offer high-quality retinal images but because the device needs to work for images captured through an ophthalmoscope which, are of much lower quality. Using such supervised algorithms offer minimal accuracy when used on images captured through an ophthalmoscope because the distribution of the testing images is entirely different
from that of the training images. Thus, an unsupervised approach is used in this project.
Most unsupervised algorithms have a segmentation step before the diameter estimation step. 
The segmentation step segments the blood vessels from the background, and this segmented vessel map is further used to estimate the diameters. Estimating the diameters can be done in various ways; one of the ways is to track the vessel edges and evaluate the shortest distance to the other parallel edge, for each point on the vessel edge. The other approach involves finding vessel centrelines and
clustering the pixels perpendicular to the centreline based on their intensities; the vessels offer
high contrast with the background and thus correspond to the low-intensity cluster. We used the latter approach.

The processing thus involves two steps:
\begin{enumerate}
\item Vessel segmentation and centreline extraction
\item Interpolation normal to the centrelines and clustering
\end{enumerate}

\subsection{Vessel segmentation and centreline extraction}
The segmentation step is crucial to locate the vessels in the image. Most segmentation algorithms
are supervised in nature and use architectures like RetinaNet trained on the DRIVE and STARE dataset. Again, these
algorithms are not suitable for images captured through an ophthalmoscope because of the low field of view and
low quality of the ophthalmoscopic images. Various unsupervised algorithms for segmentation based on match filtering, 
global thresholding, and morphological operations, etc., can be found in the literature. All the unsupervised algorithms
generally found to attain almost similar accuracies on the DRIVE and STARE dataset. We have chosen a global thresholding
based algorithm for segmentation.

\begin{figure}[H]

\begin{multicols}{3}
\includegraphics[width=0.8\linewidth]{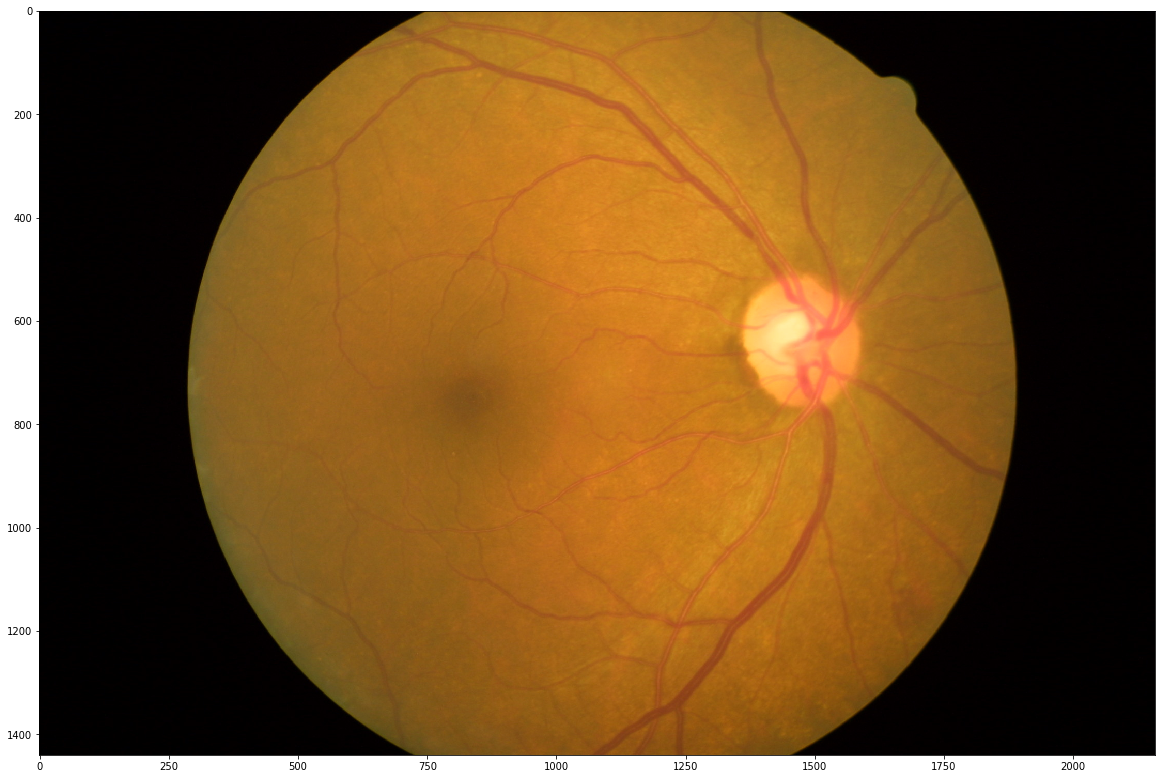}
\centering
a)\par
\includegraphics[width=0.8\linewidth]{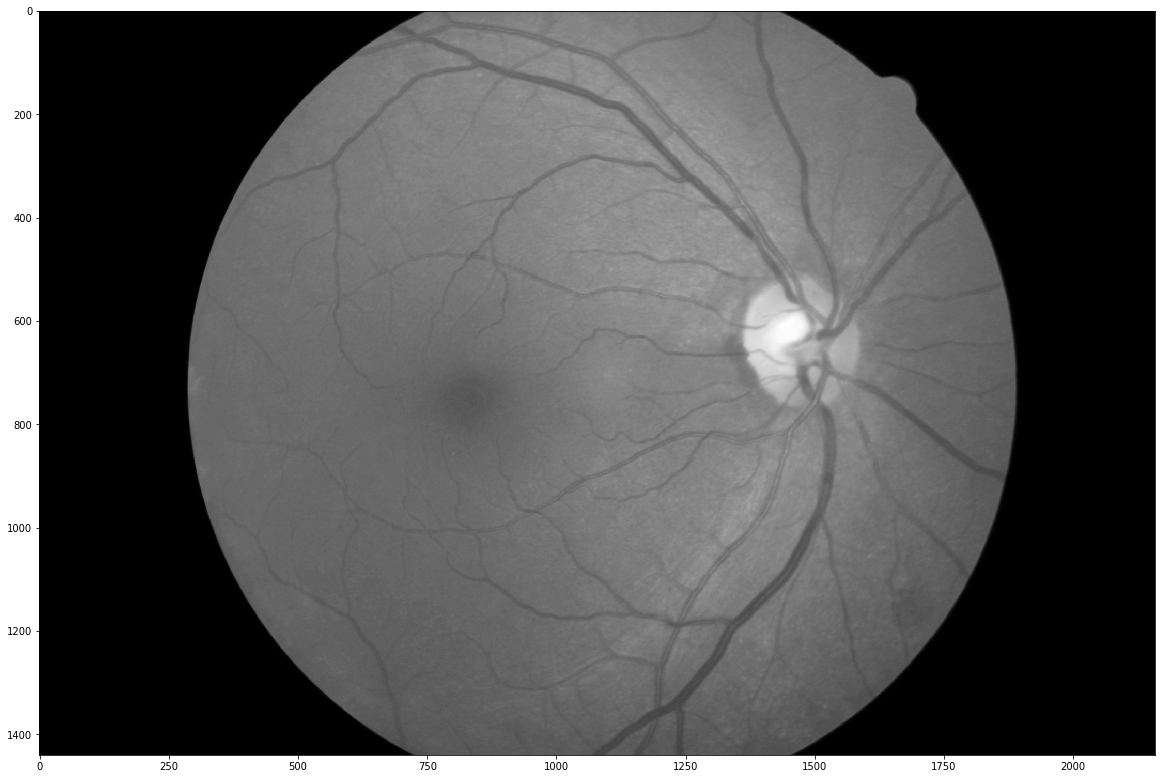}
\centering
b)\par
\includegraphics[width=0.8\linewidth]{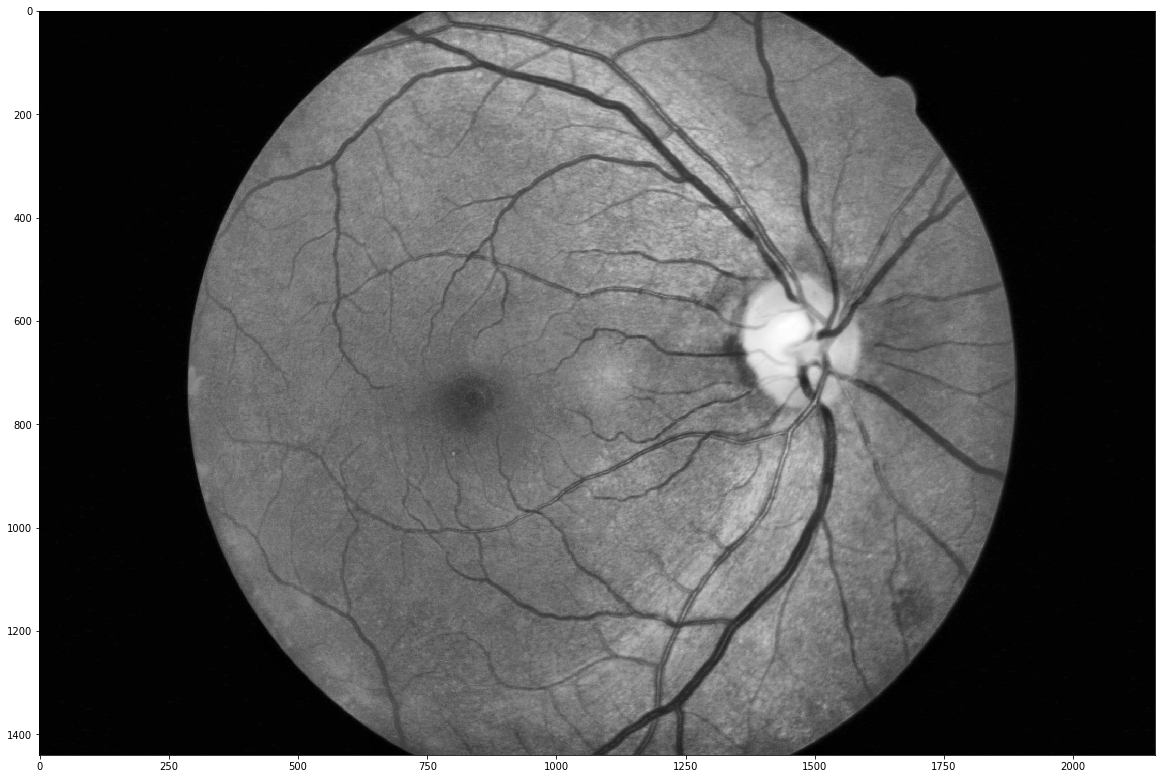}
\centering
c)\par
\end{multicols}

\begin{multicols}{3}
\includegraphics[width=0.8\linewidth]{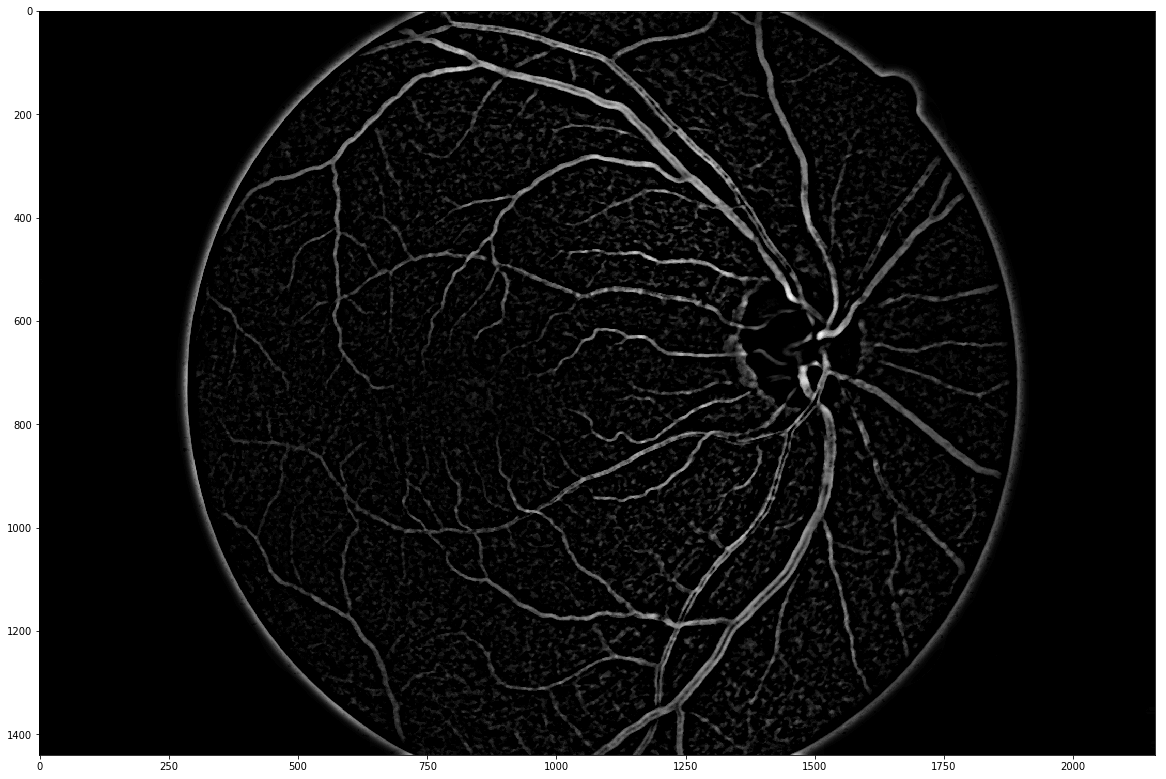}
\centering
d)\par
\includegraphics[width=0.8\linewidth]{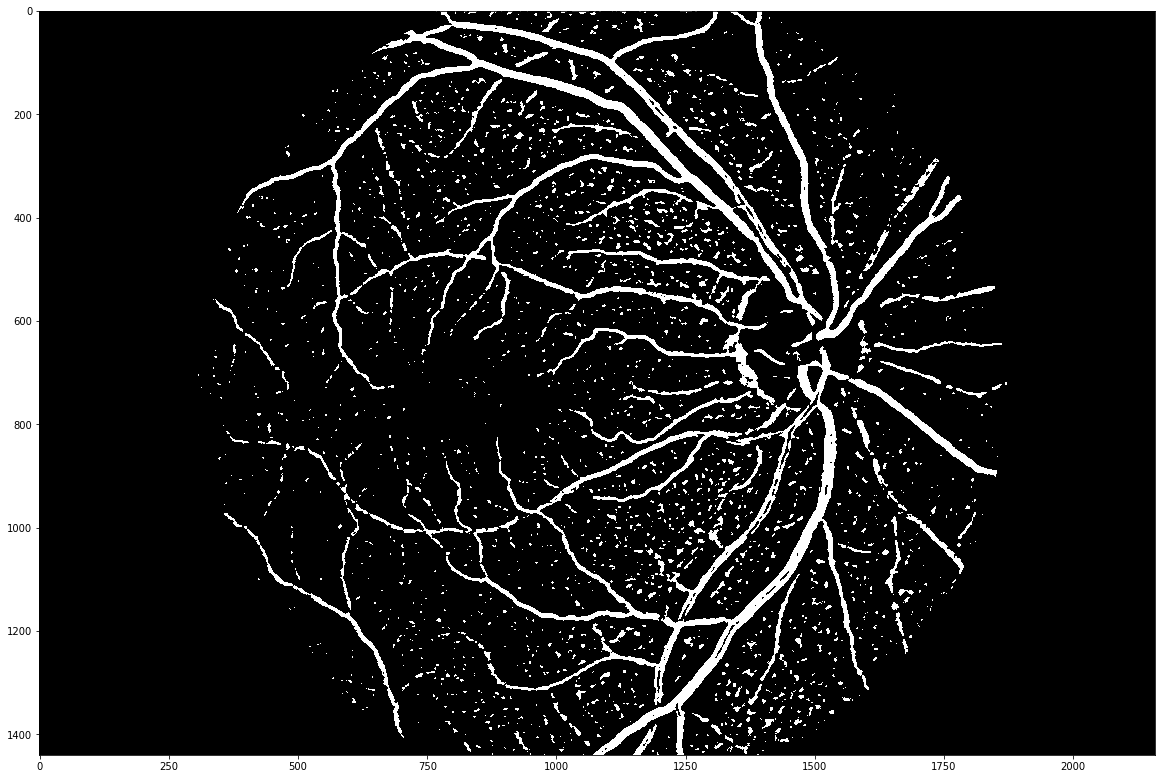}
\centering
e)\par
\includegraphics[width=0.8\linewidth]{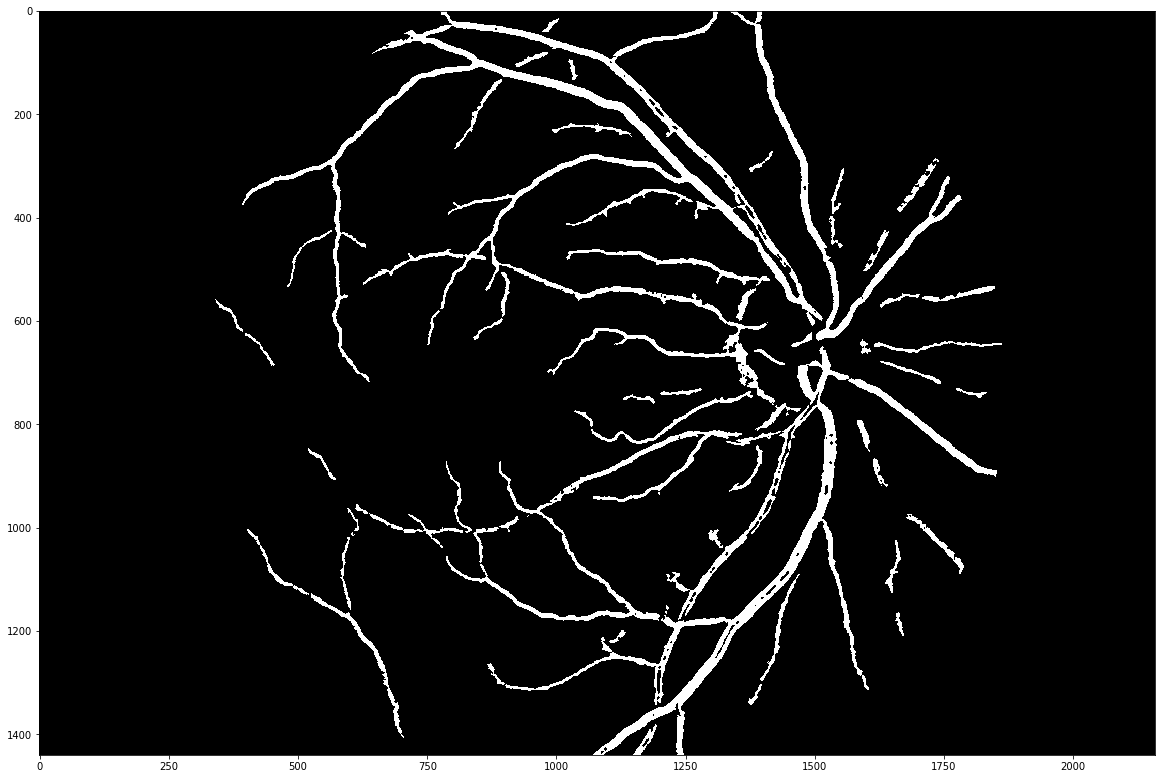}
\centering
f)\par
\end{multicols}
\caption{\textbf{Vessel segmentation:} \textbf{a)} Original image from the REVIEW dataset, \textbf{b)} Green channel of the image, \textbf{c)} Green channel enhanced using CLAHE,
\textbf{d)} Result of background subtraction, \textbf{e)} Result of global thresholding, \textbf{f)} Segmented image after noise removal}
\label{fig:segmentation}
\end{figure}

As most studies have shown that the green channel offers the most contrast with the background, the green channel
of the image was selected, and to further enhance the image Contrast Limited Adaptive Histogram Equalization (CLAHE) was applied.
To obtain a field of view(FOV) mask for the image, thresholding was used with a low threshold. To counter the boundary effects of processing the 
FOV mask was morphologically eroded using an elliptic structuring element.
The CHALE was applied using a grid of size 9x9, and the contrast clip limit value of 3.0. Background subtraction is further carried out
to extract the features from the background. The background was estimated by applying a small median filter (size=5) and a large Gaussian filter (size=55) on the enhanced image.
The enhanced image was then subtracted from the estimated background, and the negative pixels were truncated to a value of 0.
Global thresholding was further applied with a low threshold value (${\approx}$8), to obtain a rough segmented vascular map. The resulting map contained much noise in the form of false-positive blobs. This noise was removed by filling the blobs having an area less than 200px$^2$.
The result was a segmented binary map of the retinal vessels. The result of various steps on the image can be seen in Figure \ref{fig:segmentation}

\begin{figure}[H]

\begin{multicols}{5}
\includegraphics[width=0.5\linewidth]{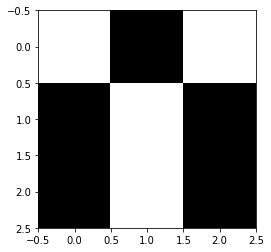}
\centering
a)\par
\includegraphics[width=0.5\linewidth]{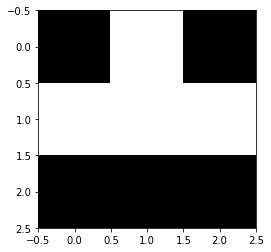}
\centering
b)\par
\includegraphics[width=0.5\linewidth]{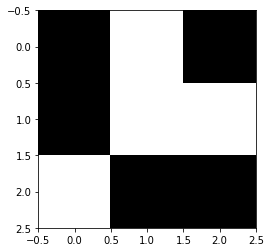}
\centering
c)\par
\includegraphics[width=0.5\linewidth]{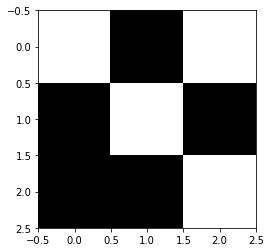}
\centering
d)\par
\includegraphics[width=0.5\linewidth]{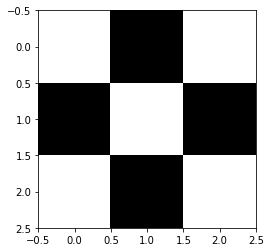}
\centering
e)\par
\end{multicols}
\caption{Tempelate kernels to detect bifurcation points}
\label{fig:kernels}
\end{figure}

\begin{figure}

\begin{multicols}{3}
\includegraphics[width=0.8\linewidth]{images/CLRIS/noise_removal.png}
\centering
a)\par
\includegraphics[width=0.8\linewidth]{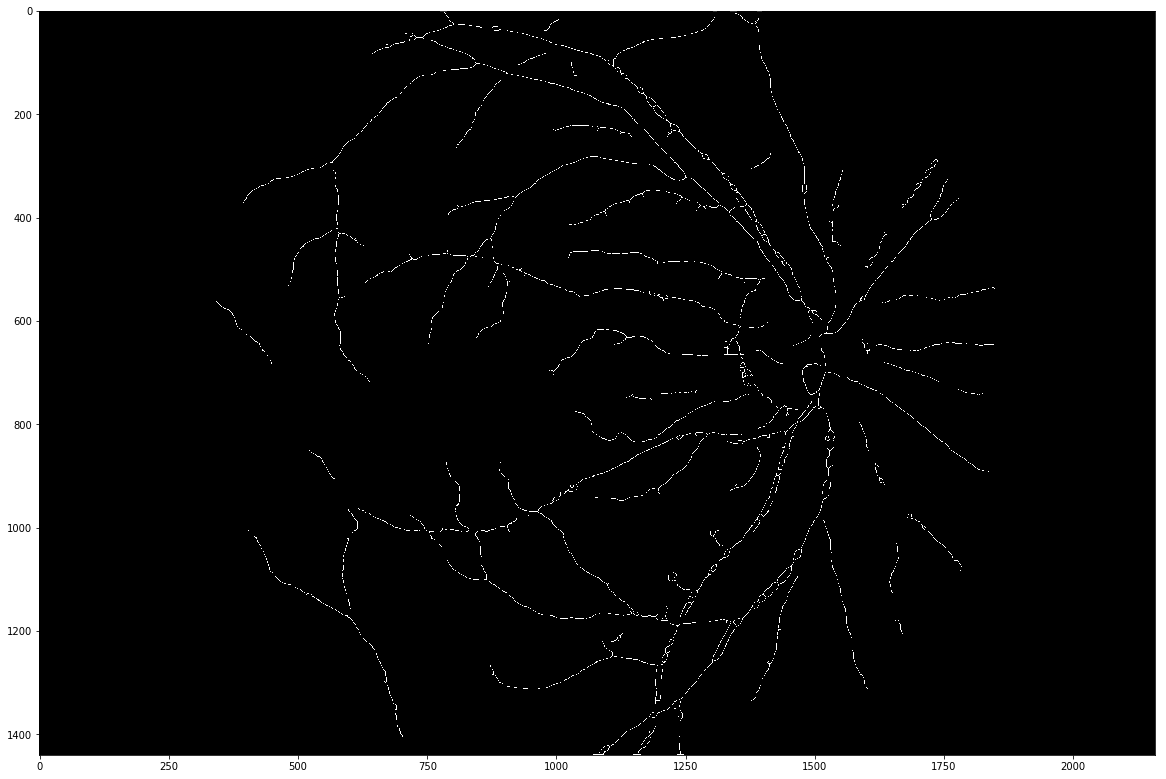}
\centering
b)\par
\includegraphics[width=0.8\linewidth]{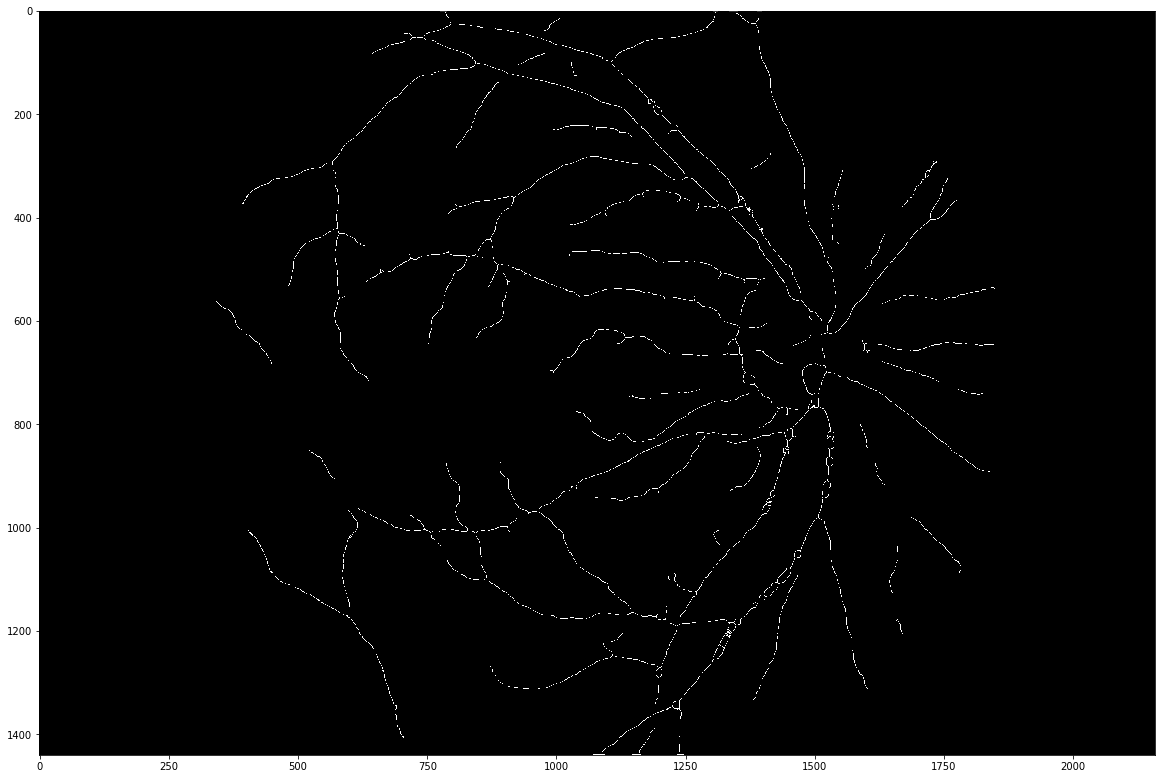}
\centering
c)\par
\end{multicols}

\begin{multicols}{3}
\includegraphics[width=0.8\linewidth]{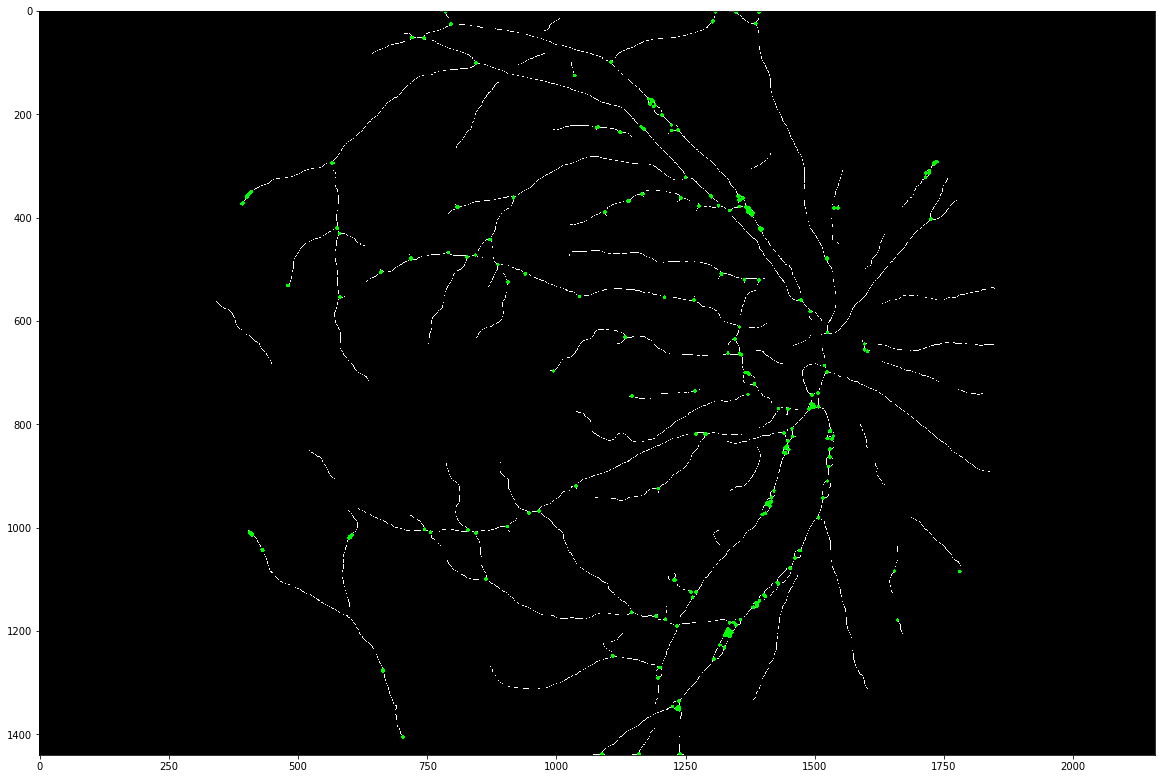}
\centering
d)\par
\includegraphics[width=0.8\linewidth]{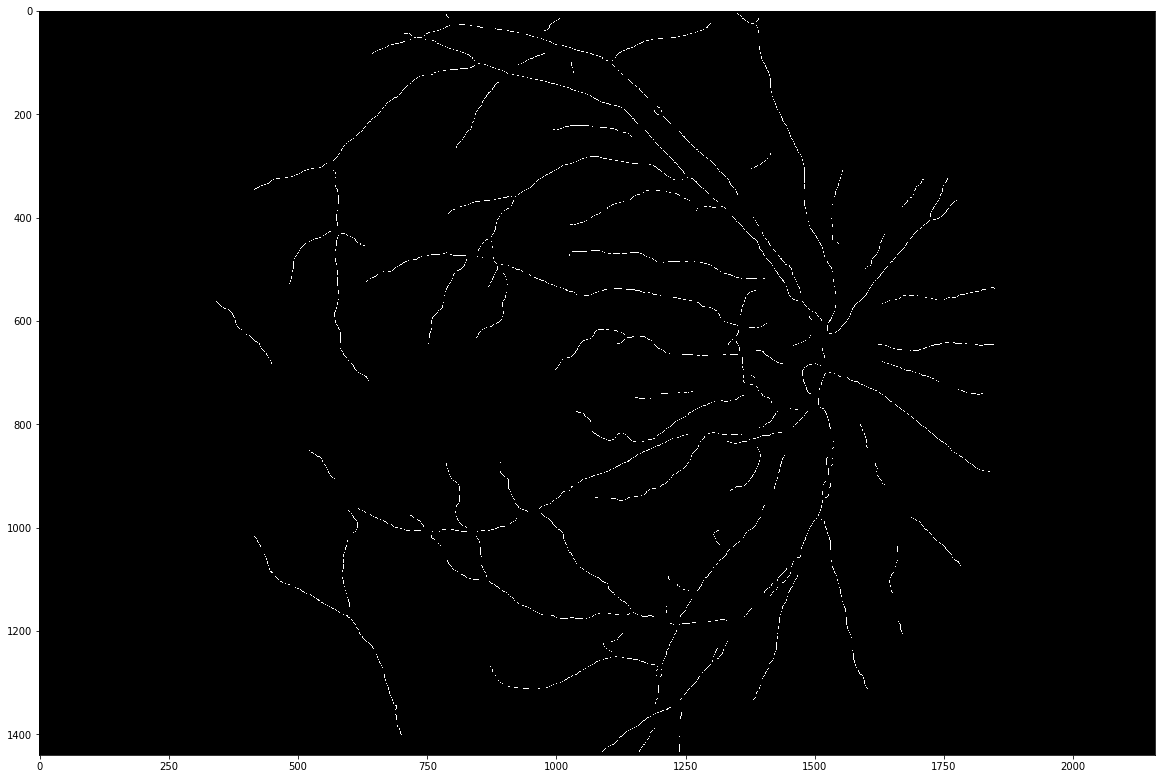}
\centering
e)\par
\includegraphics[width=0.8\linewidth]{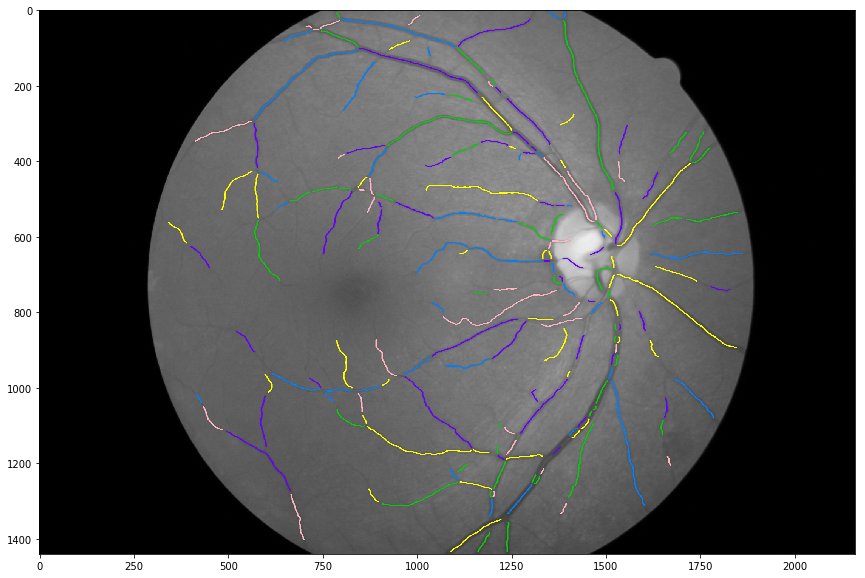}
\centering
f)\par
\end{multicols}
\caption{\textbf{Centreline extraction:} \textbf{a)} Segmented vessel map, \textbf{b)} Skeletonization using Zhang-Seun's thinning algorithm, \textbf{c)} Filled broken centrelines,
\textbf{d)} Detected bifurcation points in green, \textbf{e)} Removing bifurcation points and noise, \textbf{f)} Resuting vessel centrelines}
\label{fig:centreline}
\end{figure}

The segmented vessel map was used to estimate an upper limit for the vessel diameters, and this was done using distance transform. Distance transform gives maps each white pixel of a binary image to its least distance from the black pixels. The twice of the maximum value computed by the distance transform corresponds to the maximum vessel diameter. This estimate was very crucial for successive steps to work, as is explained further.
Once we have obtained a segmented vessel map, the centreline can be obtained using skeletonization. We have used the Zhang-Seun's thinning
algorithm to obtain the centrelines from the segmented vessel map. The morphological closing operation was used with a linear structuring element applied at every 15$^{\circ}$ for filling
broken centrelines. The resulting centreline map contains bifurcation points, and removing these points was an essential step because calculating normals at these points was ambiguous. The bifurcation points were detected using template kernels, as shown in Figure \ref{fig:kernels}, applied at every 15$^{\circ}$. The detected bifurcation points
were removed, and the resulting image was the centreline map. Due to the removal of bifurcation points, some noise was expected to be induced in the image and thus must be removed.
To remove the noise, lines that were less than 25px in length were removed. The vessel centrelines were randomly colored for better visualization. The results at each step ate shown in Figure \ref{fig:centreline}.

\subsection{Interpolation normal to the centrelines and clustering}
\begin{figure}[H]

\begin{multicols}{3}
\includegraphics[width=0.8\linewidth]{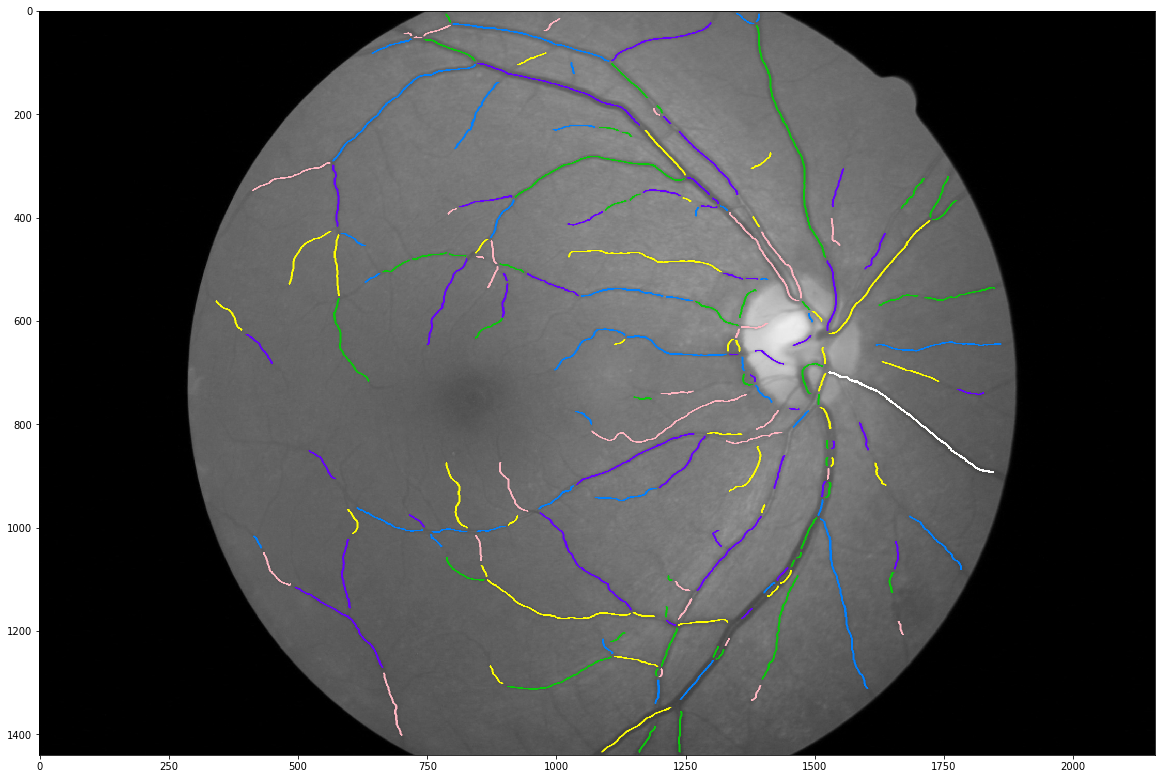}
\centering
a)\par
\includegraphics[width=0.8\linewidth]{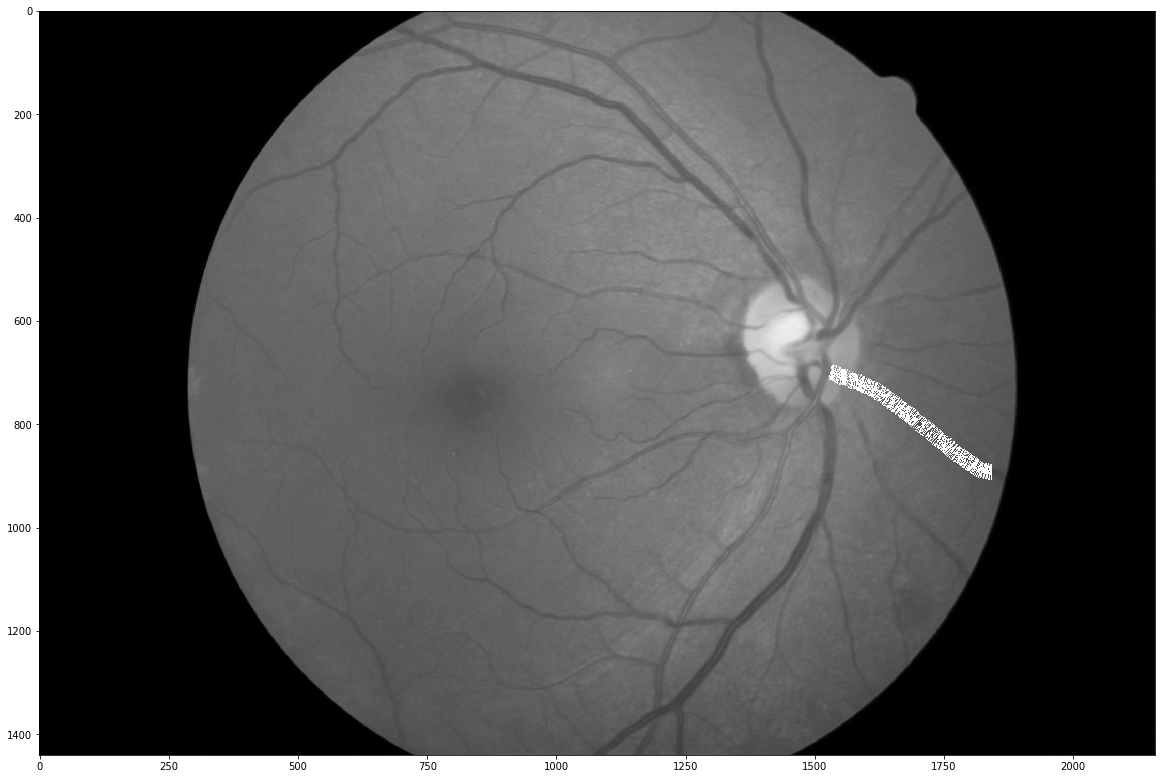}
\centering
b)\par
\includegraphics[width=0.8\linewidth]{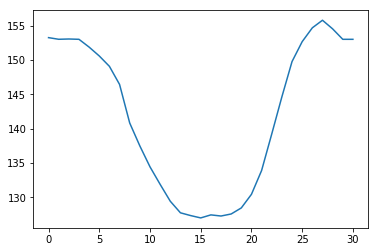}
\centering
c)\par
\end{multicols}

\begin{multicols}{2}
\includegraphics[width=0.6\linewidth]{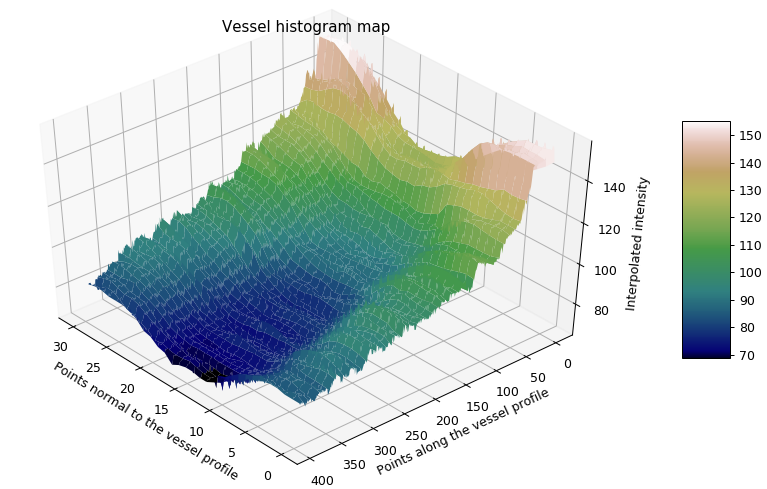}
\centering
d)\par
\includegraphics[width=0.6\linewidth]{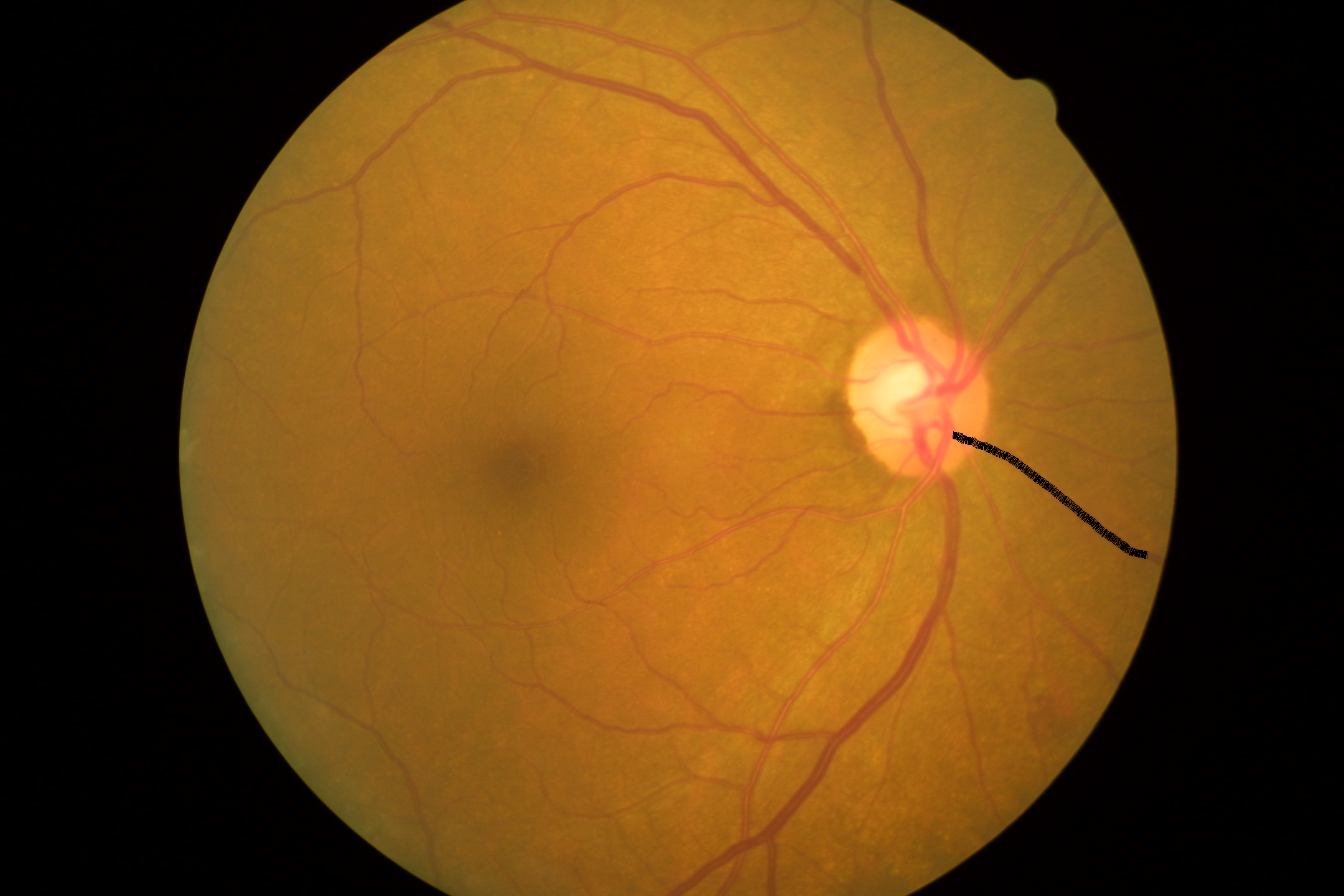}
\centering
e)\par
\end{multicols}
\caption{\textbf{Diameter estimation:} \textbf{a)} Selected vessel in white, \textbf{b)} Pixels interpolated normal to the selected vessel using the maximum diameter estimated earlier using the distance transform, \textbf{c)} Intensity variations along the normal to a particular single pixel on the selected vessel centreline,
\textbf{d)} Intensity variations along the points on the selected centreline, \textbf{e)} estimated diameters}
\label{fig:diameter}
\end{figure}

\begin{figure}[H]

\begin{multicols}{2}
\includegraphics[width=0.8\linewidth]{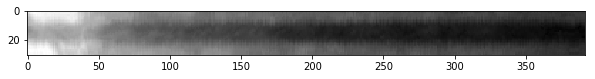}
\centering
a)\par
\includegraphics[width=0.8\linewidth]{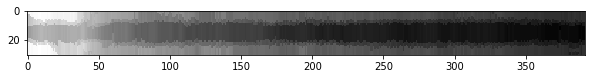}
\centering
b)\par
\end{multicols}
\centering
\includegraphics[width=0.4\linewidth]{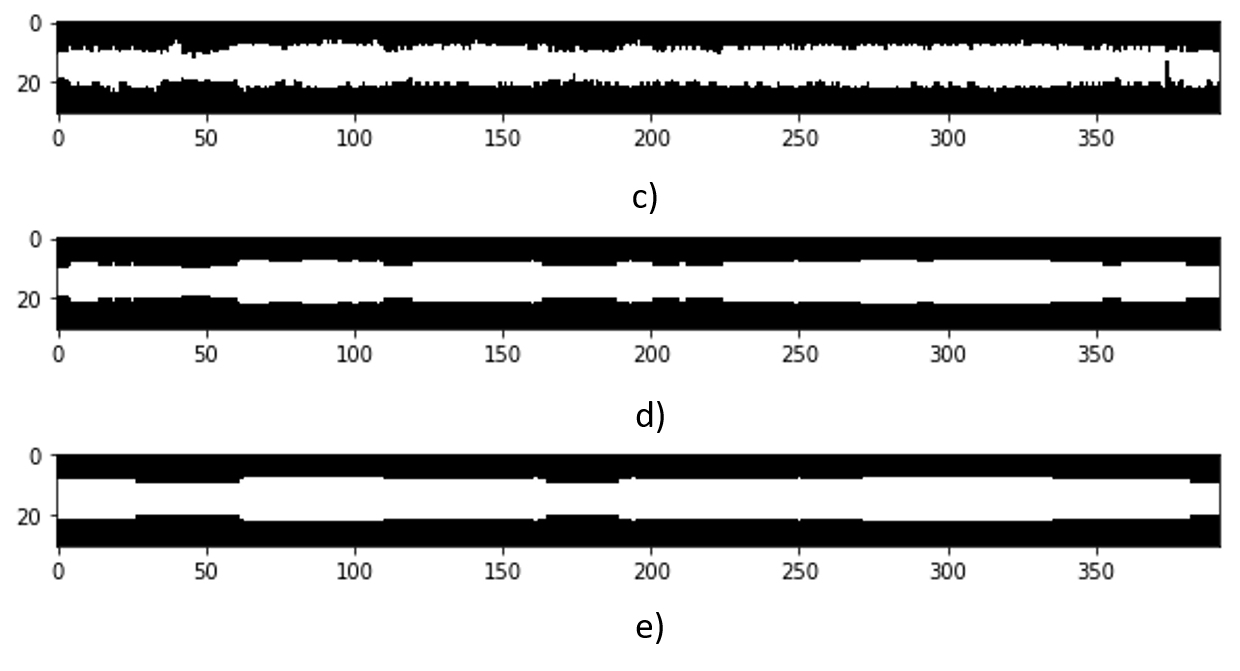}
\caption{\textbf{Diameter estimation:} \textbf{a)} Intensity variations along the normals visualized as an image, \textbf{b)} Result of clustering the intensity variation image using into 3 clusters using k-Means clustering , \textbf{c)} Result of thresholding about the lowest intensity cluster,
\textbf{d)} Exploiting symmetry about the centreline using logical and operation between the image and it's mirror image abou the centreline, \textbf{e)} Filling broken horizontal lines to smoothen the vessel profile}
\label{fig:symmetry}
\end{figure}

To select a particular vessel centreline from the centreline map, input from the user was used. The user was prompt to click on the screen. The centreline closest to the clicked point was selected. The points on the centreline were stored and normals were to be drawn at each of these points.
To draw a normal at a point, we need the slope at that point, and to calculate the slope, a window of 6 pixels centered at that point was used. The earlier computed maximum diameter estimate from the distance transform was the most crucial element in drawing the normals. Let the maximum
the estimated diameter was "d," then normals drawn were of the length "1.5xd". 
This ensured that no part of the vessel along the normal was missed. 
The points along this normal were linearly interpolated and intensity value 
 each of these points was computed using bilinear interpolation.
Thus, we can plot intensity values along any of the normal and atypical 
such variation is shown in Figure \ref{fig:diameter},c).
All such normal variations can be stacked together to obtain a 3D 
intensity plot of intensity variations aligned along a common center as 
shown in Figure \ref{fig:diameter}, d).
One can observe from the plot that the vessels correspond to the 3D 
valley. Another way to visualize these stacked-aligned intensity variations is as an image, which is shown in Figure \ref{fig:symmetry}, a).
The vascular pixels are of lower intensity in such an image. To obtain the diameters pixels in such an image can be clustered together, and the lowest intensity cluster will correspond to the vessels. Thus, K-Means clustering was used to form 3 clusters in the image, and the resulting clustered image is shown in Figure \ref{fig:symmetry}, b). To separate the vascular pixels, 
thresholding was performed about the lowest intensity cluster, and the resulting thresholded image is shown in Figure \ref{fig:symmetry}, c).

This resulting image presented a profile of the vessel aligned horizontally. However, it was noisy and had breaks. Central Light Reflux(CLR) in vessels was also a hurdle as in images with CLR; the centreline is considerably bright and, thus, usually get clustered as non-vascular.
To make our algorithm immune to CLR, we filled broken verticle lines in the profile, filling the false-negative pixels. Further, the vessel is supposed to be symmetric about the centreline. Thus, to exploit this symmetry, the image was flipped about the horizontal axis, and 'logical and operation' was performed to find the common vascular region between the flipped and the original image. The resulting image is shown in Figure 
\ref{fig:symmetry}, d).
To further smoothen the profile, horizontal points separated by a distance of less than 20pixels were joined, and the resulting final estimated vessel profile is shown in Figure \ref{fig:symmetry}, e).
To compute the diameters, the number of white pixels along each vertical line (corresponds to the earlier normals) was calculated. These computed 
diameters are shown in the Figure \ref{fig:diameter}, e).

\section{RESULTS}
The global thresholding based algorithm was tested on the DRIVE dataset and the diameter estimation algorithm on the CLRIS section of the STARE dataset. 
To test the segmentation algoritm, 20 manually annotated images were taken from the dataset. Predictions were made for each image and compared with the ground truth to find the number of true positives(TP), false positives(FP), true negatives(TN) and false negatives(FN). The metrics used to test the predictions were Pixel accuracy, sensitivity and specificity. The algorithm achieved an average pixel accuracy of 94.22\% with 69.20\% sensitivity and 96.63\% specificity. Detailed results are shown in the Table \ref{tab:DRIVE}.
\begin{equation}
  Accuracy = \frac{TP+TN}{TP+TN+FP+FN}
\end{equation}
\begin{equation}
  Sensitivity = \frac{TP}{TP+FN}
\end{equation}
\begin{equation}
  Specificity = \frac{TN}{TN+FP}
\end{equation}\\
For the diameter estimation algorithm, CLRIS section of the REVIEW dataset was used to test the algorithm. 
Predictions were made for various annotated vessel segments and the standard-deviation error (${\sigma}_{error}$) 
the metric was used to judge the algorithm. Let at a given vessel profile i, ${\chi}_{i}$ = ${\omega}_{i}$ $-$ ${\psi}_{i}$, where ${\omega}_{i}$ is the estimated 
width and ${\psi}_{i}$ is the correspondent ground truth. Let the total number of points be $n_p$ then the standard deviation of the width differences is given by

\begin{equation}
  {\sigma}_{error} = \sqrt{\frac{1}{n_p-1} \sum\limits_{1}^{n_p} {({\chi}_i-{\mu}_{error})^2}}
\end{equation}
where,
\begin{equation}
  {\mu}_{error} = \frac{1}{n_p}\sum\limits_{1}^{n_p} {\frac{1}{{\chi}_i}}
\end{equation}
The average ${\sigma}_{error}$ achieved by the algorithm is 0.40, which is in the range 1.2 - 0.2, achieved by the algorithms in the literature. Detailed results are shown in the Table \ref{tab:REVIEW}.

\begin{figure}[h]
  \includegraphics[width=\linewidth]{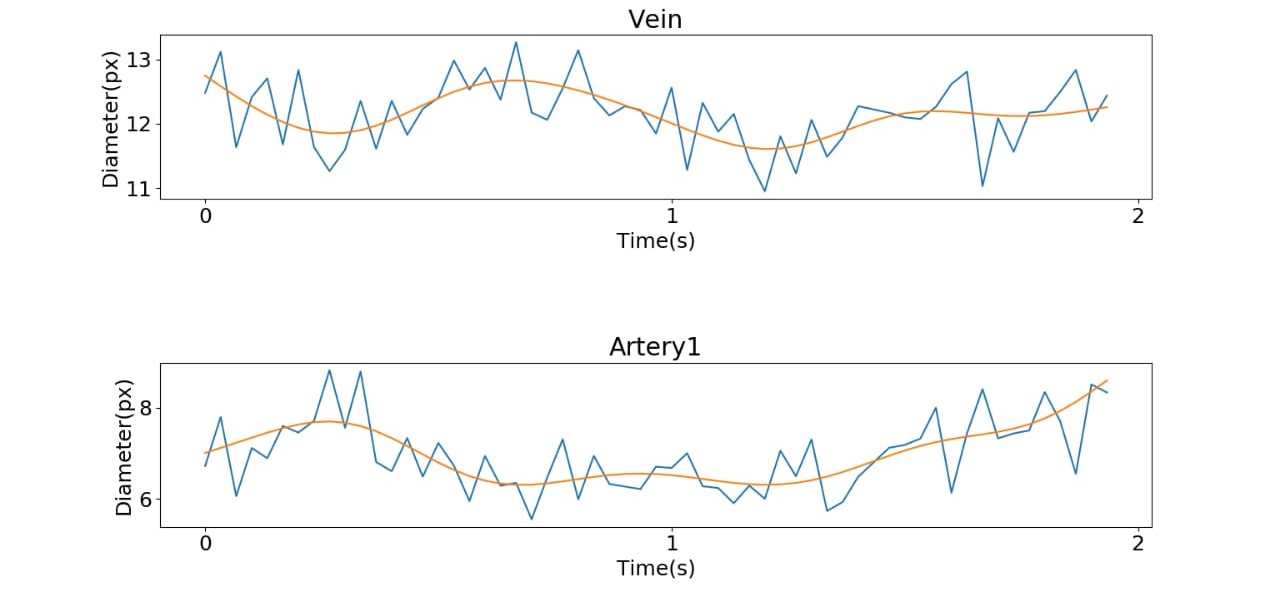}
  \caption{Time variation of vessel diameters, \color{TealBlue} Blue: original data \normalcolor  and  \color{YellowOrange} Orange: smoothened data\normalcolor}
  \label{fig:result}
  \end{figure}

A 2-second long video of fundoscopy, captured through an ophthalmoscope, was sampled at a sample rate of 30 to obtain 60 images. Each of the images was individually processed using the above algorithms. The diameters of a particular artery and a vein were tracked. The results are shown in Figure \ref{fig:result}.
The blue curve is estimated using the algorithm. The orange curve is the smoothened version of the blue curve. Savitzky-Golay filter with a window size of 0.1$\times$window-size (5) was used to smoothen the curve, and a low pass filter was applied with a cutoff frequency of 2Hz. The resulting curve clearly shows the sinusoidal variations corresponding to the SVP, as per expectations. Due to the stretchy nature of the SVP, when the arteriole diameter increase, the venule diameters are expected to decrease. Thus, the arteriole and venue diameters variation curves are expected to be a 180$^{\circ}$ out-of-phase, which is apparent in the estimated curve.

  \begin{figure}[h]
    \includegraphics[width=\linewidth]{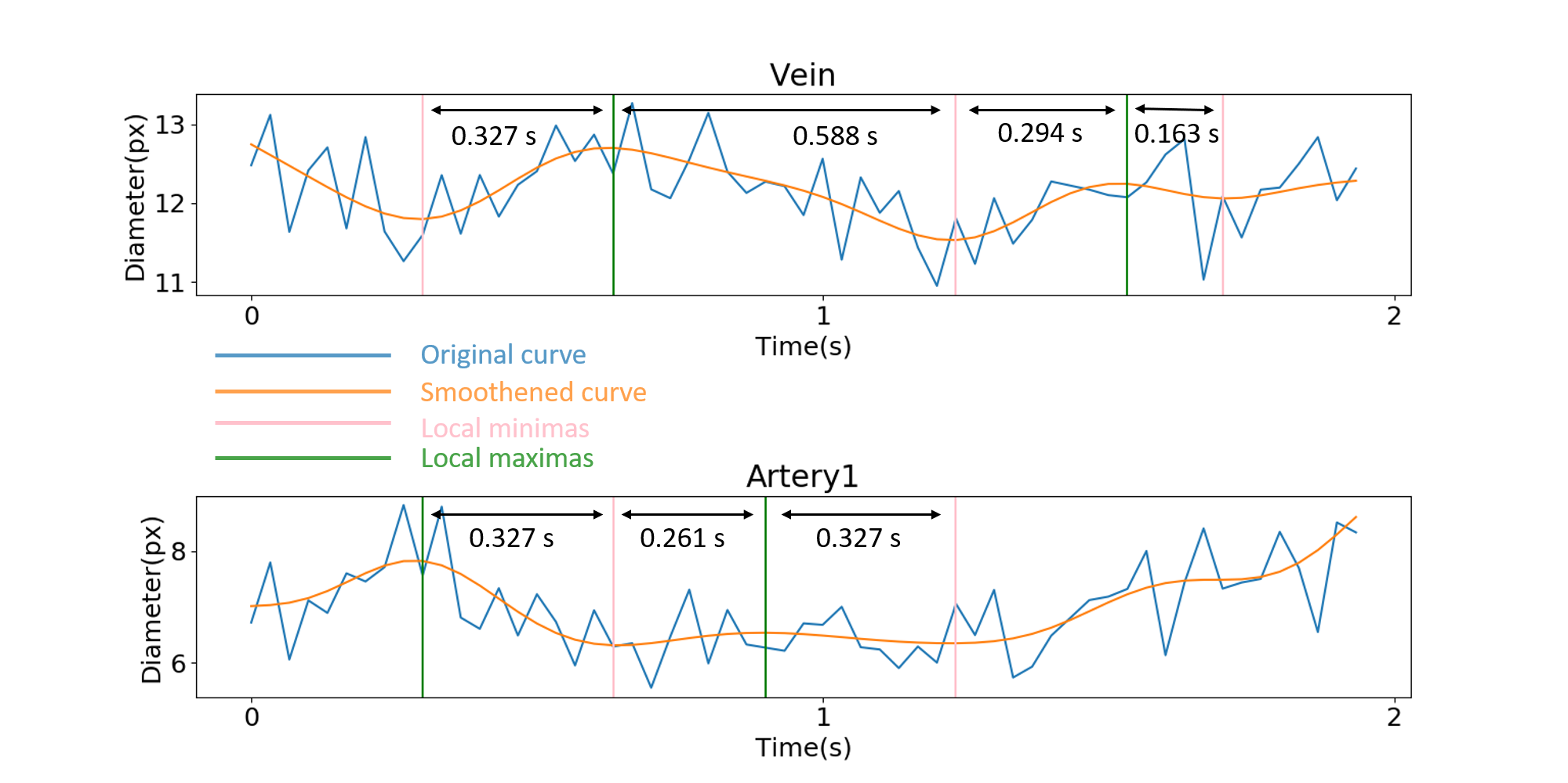}
    \caption{Estimating heart-rate by computing local maximas and minimas.}
    \label{fig:heart_rate}
    \end{figure}

\begin{table}[h]
  \caption{Performance of the segmentation algorithm on the DRIVE dataset}
  \centering
  \begin{tabular}{llll} 
  \cmidrule(r){1-4}
  Image & Accuracy  & Sensitivity & Specificity   \\
  \midrule
  Image-1 & 95.45\% & 73.25\% & 97.24\% \\ 
  Image-2 & 95.01\% & 65.36\% & 97.96\% \\ 
  Image-3 & 88.13\% & 74.71\% & 89.08\% \\ 
  Image-4 & 94.37\% & 68.61\% & 97.75\% \\ 
  Image-5 & 94.32\% & 54.47\% & 98.55\% \\ 
  Image-6 & 95.15\% & 66.75\% & 97.75\% \\ 
  Image-7 & 94.79\% & 67.40\% & 97.43\% \\ 
  Image-8 & 94.56\% & 65.58\% & 97.70\% \\ 
  Image-9 & 94.81\% & 69.22\% & 97.16\% \\ 
  Image-10 & 95.77\% & 62.89\% & 98.57\% \\ 
  Image-11 & 93.50\% & 76.17\% & 94.61\% \\ 
  Image-12 & 95.72\% & 70.89\% & 97.93\% \\ 
  Image-13 & 95.55\% & 71.74\% & 97.65\% \\ 
  Image-14 & 87.41\% & 72.51\% & 89.03\% \\ 
  Image-15 & 95.20\% & 71.17\% & 97.48\% \\ 
  Image-16 & 94.14\% & 66.59\% & 97.51\% \\ 
  Image-17 & 94.31\% & 69.32\% & 96.70\% \\ 
  Image-18 & 95.14\% & 72.87\% & 97.24\% \\ 
  Image-19 & 95.16\% & 71.15\% & 97.42\% \\ 
  Image-20 & 95.91\% & 73.27\% & 97.77\% \\ 
  \bottomrule
  Average & 94.22\% & 69.20\% & 96.63\% \\
  \end{tabular}
  \label{tab:DRIVE}
  \end{table}

To estimate heart rate from the resulting pulsation plot, first the local-maxima
and local-minima of the smoothened curve were estimated. The average
time-separation between each consecutive minima and maxima was
computed, as shown in Figure \ref{fig:heart_rate}. The estimated time-period of the pulse is then the twice of the
computed time-separation. The average time-separation for the vein came out to be \textbf{0.687 s} and for
the artery came out to be \textbf{0.6101}. The heart-rate/pulse-rate is the
number of periodic-cycles occurring in a minute (60 sec). Thus, the
heart-rate can be obtained by dividing 60 by the estimated time-periods.
The heart-rate, for the vein, came out to be \textbf{87.37}, and for the artery
came out to be \textbf{98.34}. Thus, the average heart-rate was found to be \textbf{92.85.}

\section{CONCLUSIONS}
The proposed device attached to an ophthalmoscope fulfills the need for funduscopy with in-situ vasculature analysis. The results discussed above are harmonious with the predictions. The accuracy achieved by the algorithms is comparable to that of the state-of-the-art. The efficiency of the algorithms is the topic of focus in the future. Complete automation of the device is also currently a bottleneck because the automatic alignment is susceptible to noise, and thus, the pulsations get lost in the noise, and no meaningful data is extractable. This problem can be solved by employing commercial registration programs like i2k-Retina-align, but, still, these programs cannot process the images in real-time. Thus, complete automated estimation in real-time with high efficiency is currently being worked on and will be a part of a future project.

\begin{table}[H]
  \caption{Performance of the diameter estimation algorith on the REVIEW/CLRIS dataset}
  \centering
  \begin{tabular}{llllll} 
  \cmidrule(r){1-6}
  Image & Segment  & ${\mu}_{mean}$ (px) & ${\sigma}_{mean}$ (px) & ${\mu}_{error}$ (px) & ${\sigma}_{error}$ (px) \\
  \midrule
  Image-1 & Segment-1 & 14.54 & 1.21 & -1.74 & \textbf{0.23}\\
  Image-1 & Segment-2 & 14.00 & 0.89 & -2.22 & \textbf{0.10}\\
  Image-1 & Segment-1 & 17.08 & 0.79 & -1.89 & \textbf{0.36}\\
  Image-2 & Segment-1 & 11.62 & 0.96 & 0.03 & \textbf{0.39}\\
  Image-2 & Segment-2 & 18.20 & 0.68 & -2.58 & \textbf{0.30}\\
  Image-2 & Segment-3 & 17.79 & 0.70 & -2.97 & \textbf{0.14}\\
  Image-2 & Segment-4 & 17.18 & 1.66 & -2.64 & \textbf{0.94}\\
  Image-2 & Segment-5 & 14.70 & 1.16 & -1.84 & \textbf{0.09}\\
  Image-2 & Segment-6 & 12.57 & 1.99 & 1.48 & \textbf{0.83}\\
  Image-2 & Segment-7 & 13.67 & 1.15 & 1.79 & \textbf{0.60}\\
  Image-2 & Segment-8 & 12.00 & 1.41 & 0.28 & \textbf{0.43}\\
  Image-2 & Segment-9 & 12.20 & 0.84 & -1.11 & \textbf{0.78}\\
  Image-2 & Segment-10 & 10.5 & 0.93 & 0.49 & \textbf{0.19}\\
  Image-2 & Segment-11 & 11.09 & 2.81 & -2.26 & \textbf{0.60}\\
  Image-2 & Segment-12 & 14.10 & 1.66 & 0.73 & \textbf{0.11}\\
  Image-2 & Segment-13 & 14.44 & 0.70 & -1.90 & \textbf{0.20}\\
  Image-2 & Segment-14 & 11.32 & 1.42 & -0.48 & \textbf{0.22}\\
  Image-2 & Segment-15 & 8.29 & 0.96 & 0.07 & \textbf{0.13}\\
  Image-2 & Segment-16 & 17.43 & 4.59 & 4.07 & \textbf{1.00}\\
  Image-2 & Segment-17 & 8.31 & 2.46 & 0.56 & \textbf{0.12}\\
  Image-2 & Segment-18 & 7.00 & 1.75 & -0.76 & \textbf{0.20}\\
  \bottomrule
  Average & & & & & \textbf{0.40}\\
  \end{tabular}
  \label{tab:REVIEW}
  \end{table}

\bibliographystyle{unsrt}  


\end{document}